\documentclass[a4paper]{spie}

\usepackage[colorlinks=true, allcolors=blue]{hyperref}
\usepackage[top=2.54cm, bottom=4.94cm, left=1.925cm, right=1.925cm]{geometry}
\usepackage{graphics}
\usepackage{graphicx}
\usepackage{bm}
\usepackage{amsmath}
\usepackage{amsfonts}
\usepackage{amssymb}
\usepackage{latexsym}
\usepackage[dvipsnames]{xcolor}
\usepackage{bbold}
\usepackage{braket}

\usepackage{tikz,tikz-3dplot,bm}
\usepackage{hyperref}
\usepackage{physics}
\usepackage{float}
\begin{document}
	
\author[a]{Benjamin Dawson}
\author[b]{Nicholas Furtak-Wells}
\author[a]{Thomas Mann}
\author[a]{Gin Jose}
\author[b]{Almut Beige}
\affil[a]{School of Chemical and Process Engineering, University of Leeds, Leeds LS2 9JT, United Kingdom}
\affil[b]{School of Physics and Astronomy, University of Leeds, Leeds LS2 9JT, United Kingdom}

\authorinfo{Further author information: (Send correspondence to Benjamin Dawson)\\Benjamin Dawson: E-mail: py13bhd@leeds.ac.uk}	
	
\title{Spontaneous emission of atomic dipoles near two-sided semi-transparent mirrors}

\maketitle

\begin{abstract}
Atom-field interactions near optical interfaces have a wide range of applications in quantum technology. Motivated by this, this paper revisits the spontaneous emission of atomic dipoles in the presence of a two sided semi-transparent mirror. First we review the main properties of the quantised electromagnetic field near a semi-transparent mirror. To do so, we employ a quantum mirror image detector method which maps the experimental setup which we consider here onto analogous free space scenarios. We emphasise that the local density of states of the electromagnetic field depends on the reflection rates of both sides of the mirror surface. Hence it is not surprising that also the spontaneous decay rate of an atomic dipole in front of a semi-transparent mirror depends on both reflectance rates. Although the effect which we describe here only holds for relatively short atom-mirror distances, it can aid the design of novel photonics devices.
\end{abstract}

\date{\today}

\section{Introduction}

\begin{figure} [t]
	\begin{center}
		\begin{tabular}{c} 
			\includegraphics[width=8cm]{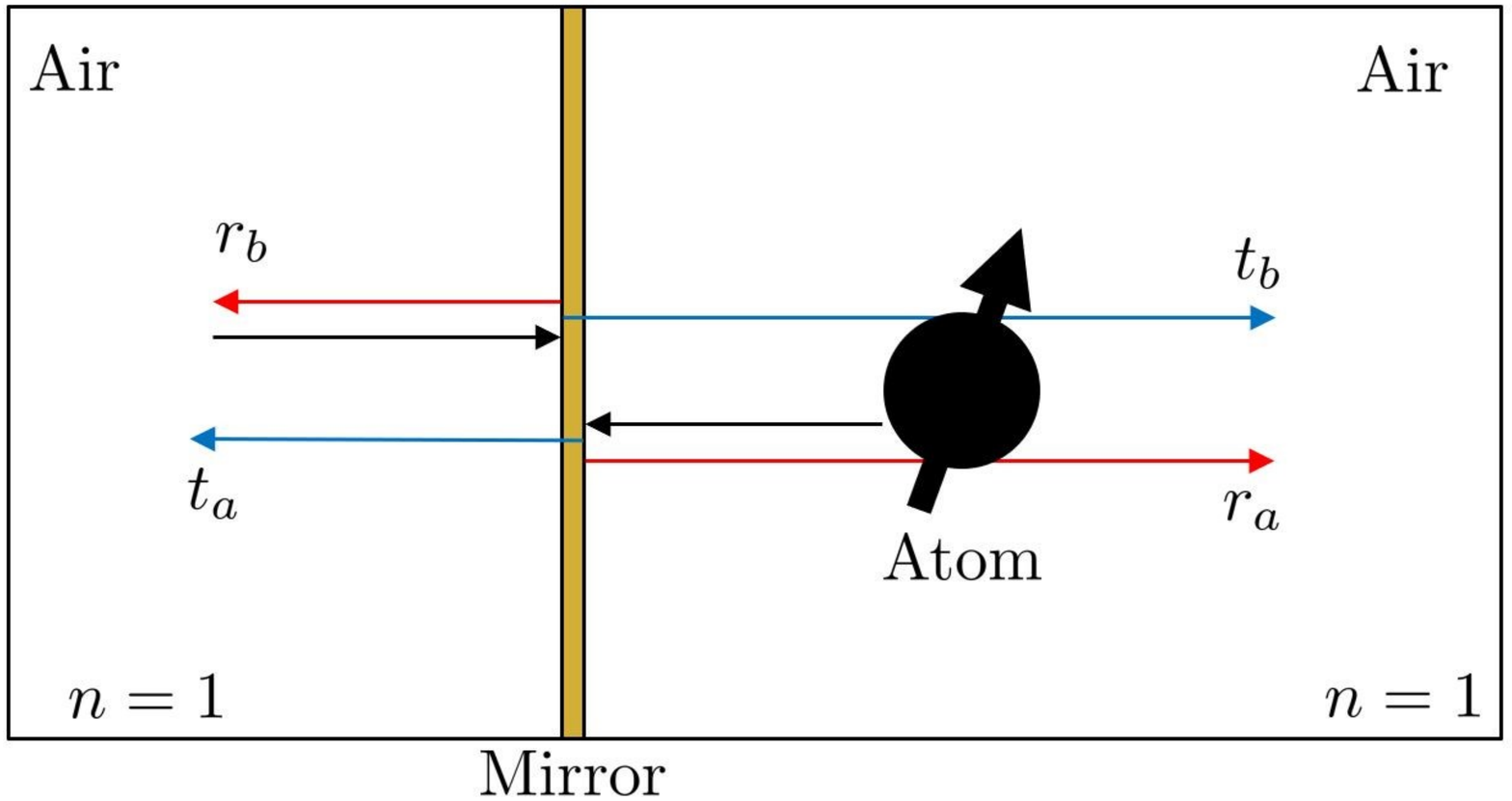}
		\end{tabular}
	\end{center}
	\caption[figpaperlogo] 
	{ \label{figpaperlogo} 
		Schematic view of the experimental setup which we consider in this paper. It contains a fluorescent atom in front if a semi-transparent mirror. For simplicity we assume that the medium on both sides of the mirror is air which has the refractive index $n=1$. Light emitted from the atom is both transmitted though and reflected back from the interface with rates $t_a$ and $r_a$ respectively. Light incident from the opposite side is transmitted and reflected at rates $t_b$ and $r_b$.}
\end{figure}

Recent technological developments in silicon photonics encourage the design of novel devices with a wide range of applications in quantum technology \cite{thrush,zhang, xu}. For example, devices which utilize the sensitivity of an atomic dipole's fluorescent properties to its respective environment have potential applications in quantum sensing \cite{tuchin}. Motivated by these developments, this paper reviews a quantum mirror image detector method which can be used to model the quantised electromagnetic field in the presence of optical interfaces for a wide range of experimental parameters \cite{furtak,dawson}. The experimental set up which we consider here is shown in Fig.~\ref{figpaperlogo} and consists of an atomic dipole placed near a two-sided semi-transparent mirror. In the following, we describe the mirror surface by its reflection and transmission rates $r_a$, $r_b$, $t_a$ and $t_b$ with 
\begin{eqnarray} \label{1}
r_a^2 + t_a^2 = r_b^2 + t_b^2 &=& 1 
\end{eqnarray}
with indices $a$ and $b$ referring to wave packets approaching the mirror form the left and from the right, respectively. For simplicity, we neglect the possible absorption of light in the mirror surface, the presence of evanescent field modes as well as any dependence of the reflection and transmission rates on angle, frequency and polarization of the incoming light.

The modelling of atom-field interactions near highly-reflecting mirrors and dielectric media already attracted a lot of attention in the literature (see e.g.~Refs.~\cite{morawitz, carniglia, Agarwal, wylie, Dalton,in,in2, Vogel, Glauber, Barnett, khosravi, zakowicz, creatore, furtak, jake, dawson,drabe,kuhn}). The most common method of modelling the electromagnetic field in the presence of a semi-transparent mirror is the triplet-mode model by Carniglia and Mandel \cite{carniglia}. Its basic idea is to assume that the electromagnetic field contains only stationary energy quanta with accordingly-weighted incident, reflected and transmitted components. When using this approach to calculate atomic decay rates in the presence of an optical interface, predictions closely match experimental data (see e.g.~Refs.~\cite{drexhage, snoeks, amos, blatt}). However, when modelling light incident onto a semi-transparent mirror from both sides, the triplet-mode model leads to unphysical interference effects \cite{zakowicz, creatore}. Light reflection by optical mirrors still needs to be studied in more detail \cite{jake}. 

There are different ways of avoiding the above mentioned interference problem. For example, one can adopt the input-output formalism \cite{in,in2}, so-called universe mode models \cite{Dalton} or assume that the electromagnetic field contains positive as well as negative frequency components \cite{khosravi,creatore,jake}.  Unfortunately, the  consistency of these different approaches has not yet been shown. In this paper we therefore adopt an alternative approach\textemdash a quantum mirror image detector method \cite{furtak,dawson}. Inspired by the method of mirror images in classical electrodynamics, this method maps the dynamics of wave packets scattering from a semi-transparent mirror onto analogous free space scenarios, thereby avoiding the question of how to describe the mirror surface. We then use this approach to calculate the spontanous decay rate of an atomic dipole in the presence of a two-sided semi-transparent mirror. As we see below, this rate depends on both reflection rates, $r_a$ and $r_b$, of the mirror interface.

This paper comprises of four sections. In Section \ref{sec2} we review the main results of Refs.~\cite{furtak,dawson} and describe the main assumptions made in the derivation of the electric field observable for the experimental setup which is shown in Fig.~\ref{figpaperlogo}. Section \ref{sec3} presents an expression for the corresponding atomic decay rate and illustrates its dependence on the atom-mirror distance and other mirror parameters. Lastly, Section \ref{secconclusions} contains a summary of our findings and discusses their implications.

\section{The quantum mirror image detector method} \label{sec2}

To model the experimental setup in Fig.~\ref{figpaperlogo}, we map the dynamics of incoming wave packets onto the dynamics of wave packets in analogous free space scenarios \cite{furtak,dawson}. To do so, we distinguish between wave packets originating on the left hand side and wave packets originating on the right hand side of the mirror interface. Placing the mirror surface at $x=0$, both cases correspond to $x \ge 0$ and $x < 0$, respectively. In the following, we label the corresponding electric field contributions by superscripts $(a)$ and $(b)$. Moreover we assume that wave packets travel as they do in free space, i.e.~as predicted by Maxwell's equations in the absence of any reflecting surfaces.  

In the presence of the semi-transparent mirror shown in Fig.~\ref{figpaperlogo}, the transmitted part of an incoming wave packet propagates exactly as it would in free space. However, the reflected part will eventually be found at a position $-x$ instead of arriving at the position $x$ of the transmitted field. According to the quantum mirror image detector method, the general solution of the classical electric field in the presence of a semi-transparent mirror ${\mathbf E}_{\text{mirr}}(\mathbf r,t)$ at a position ${\bf r}$ and at a time $t$ is therefore of the form \cite{furtak,dawson} 
\begin{eqnarray}\label{Efieldinterface}
{\mathbf E}_{\text{mirr}}(\mathbf r,t) &=& \left[ {\bf E}^{(b)}_{\rm free}({\mathbf {r}}, t) + r_b \, {\bf E}^{(b)}_{\rm free}(\tilde{{\mathbf {r}}}, t,\phi _3) + t_a \, {\bf E}^{(a)}_{\rm free}({\mathbf {r}}, t,\phi_4) \right] \Theta (-x) \nonumber \\
&& + \left[ {\bf E}^{(a)}_{\rm free}(\mathbf r, t) + r_a \, {\bf E}^{(a)}_{\rm free}(\tilde{\mathbf{r}}, t,\phi _1) + t_b \, {\bf E}^{(b)}_{\rm free}(\mathbf r, t,\phi _2) \right] \Theta(x) 
\end{eqnarray}
with $\tilde{{\mathbf {r}}}=(-x,y,z)$ and with the Heaviside step function $\Theta (x)$ given by
\begin{eqnarray}
\Theta (x) &=& \left\{ \begin{array}{ll} 1 &  x\geq 0 \, , \\ 0 & x<0 \, . \end{array} \right. 
\end{eqnarray}
Moreover, ${\bf E}^{(i)}_{\rm free}({\mathbf {r}}, t,\phi_j)$ with $i = a,b$ and $j=1,\ldots,4$ is a free space solution of Maxwell's equations with the same initial conditions as ${\mathbf E}_{\text{mirr}}(\mathbf r,t)$. The superscripts $^{(a)}$ and $^{(b)}$ indicate whether an incoming wavepacket approaches the mirror from the right or from the left, respectively. Finally, the phase factors $\phi_j$ in Eq.~(\ref{Efieldinterface}) are free parameters which can be used to model phase factor changes of incoming wave packets upon reflection and transmission. Note that these phase factors need to obey the condition
\begin{equation}
\phi_1 - \phi_2 + \phi_3-\phi_4 = \pm (2n +1)\pi \, , 
\end{equation} 
where $n$ is an integer, to ensure that energy is preserved \cite{furtak,dawson}.

Modelling the experimental setup in Fig.~\ref{figpaperlogo} from a quantum optics perspective is made difficult by the fact that its electromagnetic field is not a closed system and does not evolve independent of the mirror surface \cite{furtak, dawson}. However, demanding that electric and magnetic field expectation values evolve exactly as predicted by classical electrodynamics, one can show that the observable ${\bf E}_{\text{mirr}}(\mathbf r)$ of the electric field at a position ${\bf r}$ need to be of the form \cite{furtak,dawson}
\begin{eqnarray}\label{normalisedEmirr}
{\bf E}_{\text{mirr}}(\mathbf r) &=& \left[ {1 \over \eta_b} \, {\bf E}^{(b)}_{\rm free} (\mathbf r) + {r_b \over \eta_b} \, {\bf E}^{(b)}_{\rm free}( \tilde{\mathbf r} , \phi _3) + {t_a \over \eta_a} \, {\bf E}^{(a)}_{\rm free}( \mathbf r , \phi _4) \right] \Theta (-x) \nonumber \\
&& + \left[ {1 \over \eta_a} \, {\bf E}^{(a)}_{\rm free}(\mathbf r) + {r_a \over \eta_a} \, {\bf E}^{(a)}_{\rm free}(\tilde{\mathbf r},\phi _1) + {t_b \over \eta_b} \, {\bf E}^{(b)}_{\rm free}( \mathbf r , \phi _2) \right] \Theta(x) \, ,
\end{eqnarray}
where ${\bf E}^{(i)}_{\rm free} (\mathbf r)$ denote the observable of the electric field in free space. As before, the superscripts $(a)$ and $(b)$ indicate whether a wave packet originates from the right or from the left half space of the mirror, respectively. Since the quantum mirror image detector method requires that we measure the photons incident on either side of the mirror differently means our mapping onto analogous free space scenarios is only possible, if we effectively double the Hilbert space of the electromagnetic field. Other papers also emphasise the need for such a doubling \cite{khosravi,creatore,jake}. 

\begin{figure}[t]
	\begin{center}
		\begin{tabular}{c}
			\includegraphics[width=0.47 \textwidth]{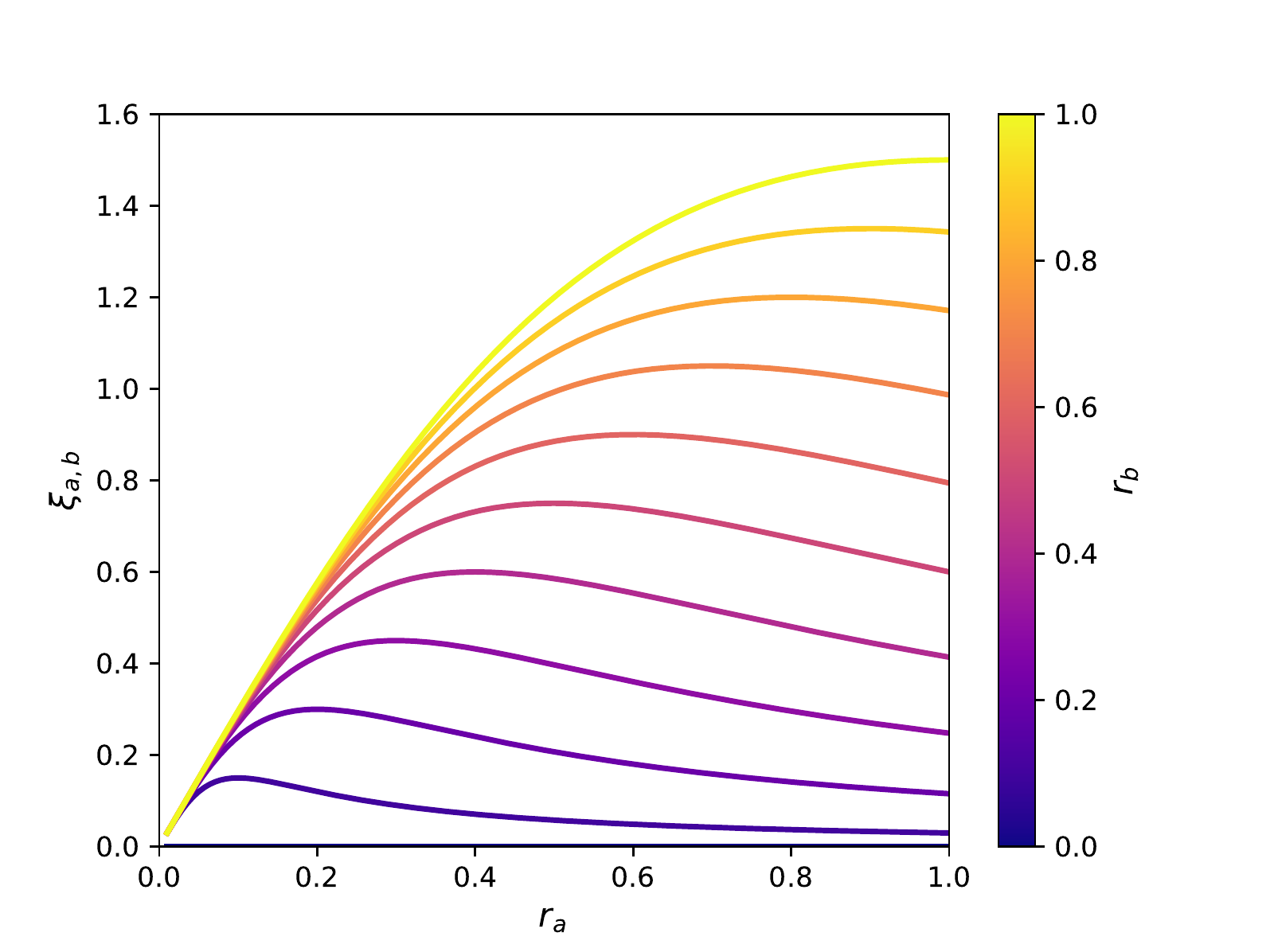}
		\end{tabular}
	\end{center}
	\caption[figxi] 
	{\label{figxi}
		For non-absorbing, energy-conserving mirror surfaces the mirror parameter $\xi_{a,b}$ varies between $0$ for free space and $1.5$ for a perfect mirror.}
\end{figure}

The parameters $\eta_a$ and $\eta_b$ in Eq.~(\ref{normalisedEmirr}) are normalisation factors. Unfortunately, these cannot be determined, as usual, by simply assuming that a photon of frequency $\omega$ has the energy $\hbar \omega$. The reason for this is that the energy of the system resides not only in the electromagnetic field. Some of it is contained in the mirror surface (for more details see Ref.~\cite{furtak}). Currently, these are calculated by demanding that the spontaneous decay rate $\Gamma_{\text{mirr}}(x)$ of an atom interacting with the electric field in Eq.~(\ref{normalisedEmirr}) reduces for relatively large atom-mirror distances $x$ to its respective free space value $\Gamma_{\rm free}$.
From this we get 
\begin{eqnarray}
\frac{1+r_a^2}{\eta_a^2} + \frac{t_b^2}{\eta_b^2} = \frac{1 + r_b^2}{\eta^2_b} + \frac{t_a^2}{\eta^2_a} &=& 1 \, ,
\end{eqnarray}
and following the procedure given in Refs. \cite{furtak,dawson},
we find that the normalisation factors $\eta_a$ and $\eta_b$ are given by
\begin{eqnarray}
\eta^2_{a} = 1 + \frac{r_a^2}{r_b^2} ~~ {\rm and} ~~ \eta^2_{b} = 1 + \frac{r_b^2}{r_a^2} 
\label{normalisationconsts}
\end{eqnarray}
providing the mirror surface does not absorb energy from the electromagnetic field.
As a result of such a locality condition, the local density of states of the electromagnetic field in the immediate vicinity of the mirror surface differs in general from what it would be in the absence of an interface  \cite{furtak,dawson}. Near the $x=0$ plane, it depends on the reflection rates $r_a$ and $r_b$ of both sides of the mirror interface. 

\section{Atomic spontaneous decay rates} \label{sec3}

\begin{figure}[t]
\begin{center}
	\begin{tabular}{c} 
		\includegraphics[width=0.47 \textwidth]{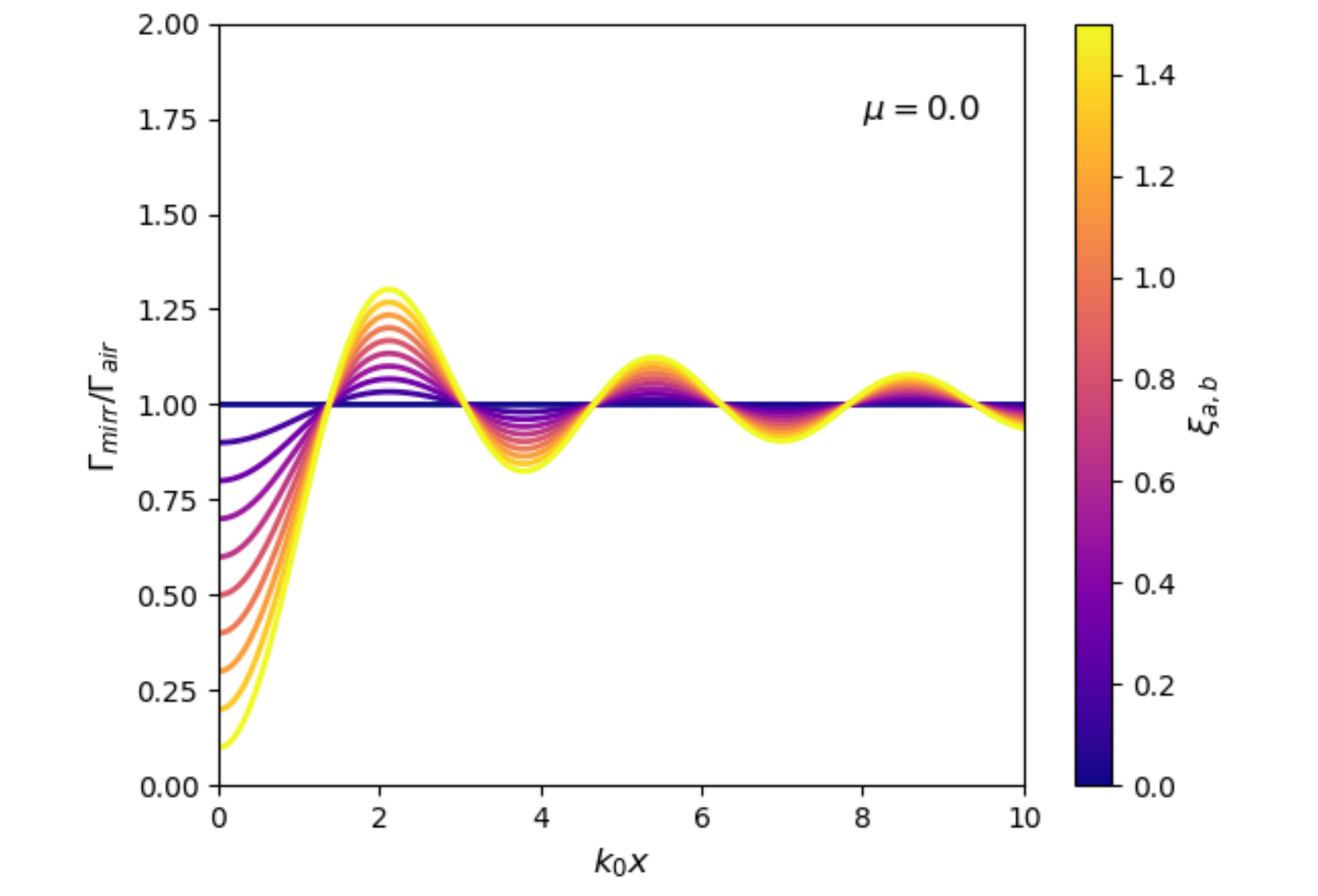}
		\includegraphics[width=0.47 \textwidth]{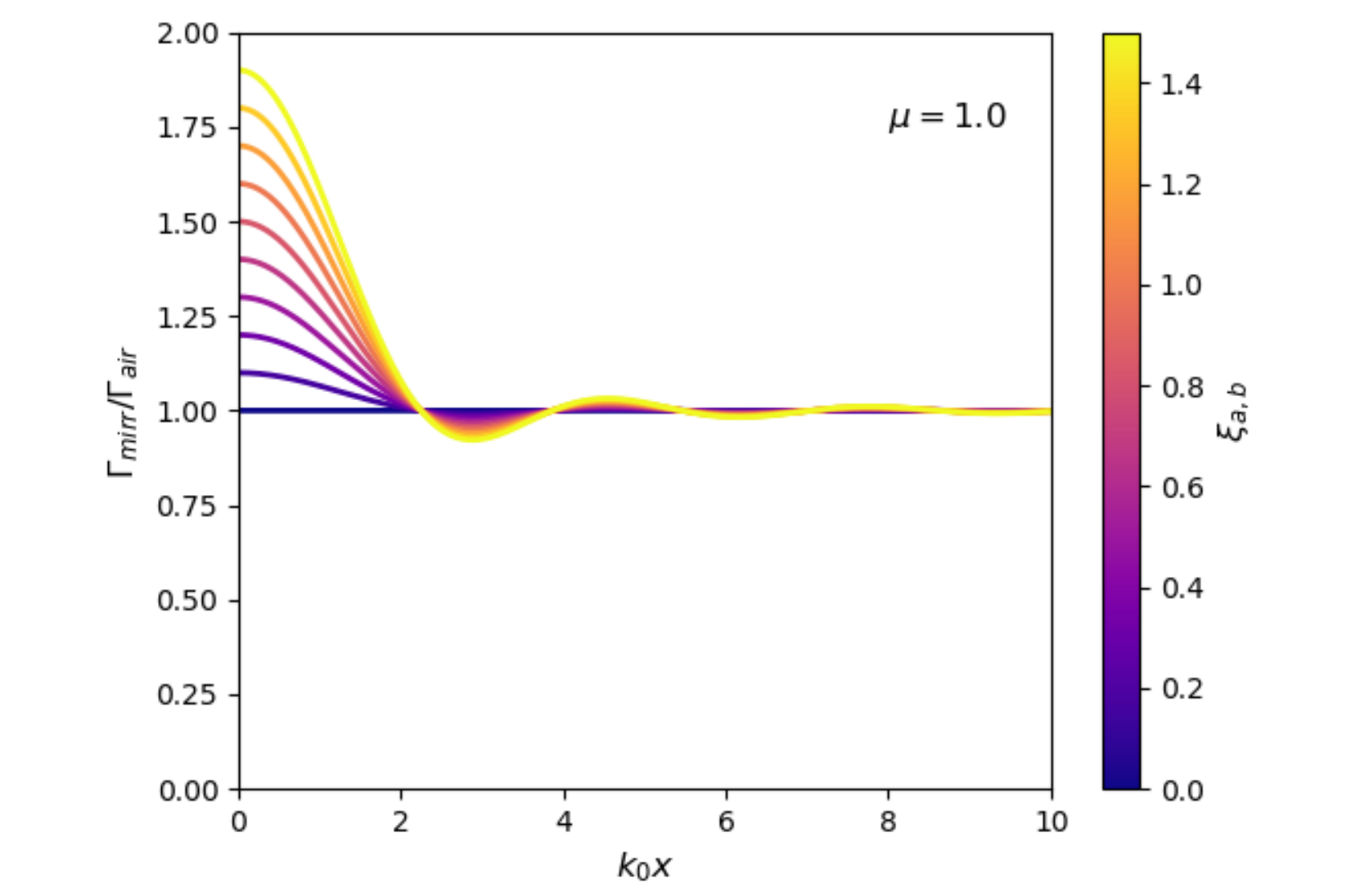}
	\end{tabular}
\end{center}
\caption[figmudep5] 
{ \label{figmudep5} 
	The dependence of the spontaneous decay rate $\Gamma_{\rm mirr}(x)$ on the mirror parameter $\xi_{a,b}$ and distance from the mirror surface positioned at $x=0$. }
\end{figure} 

Proceeding as described in Refs.~\cite{furtak,dawson}, the electric field observable in Eq.~(\ref{normalisedEmirr}) can be used to show that the spontaneous decay rate $\Gamma_{\text{mirr}}(x)$ in the presence of a two-sided semi-transparent mirror equals 
\begin{eqnarray}\label{decayratesimplified} 
\Gamma_{\text{mirr}}(x) &=& \left[1 - \xi_{a,b} \, \biggr( \frac{\cos(2k x)}{(2k x)^2} - \frac{\sin(2k x)}{(2k x)^3}\biggr)(1+\mu) - \xi_{a,b} \, \frac{\sin(2k x)}{2k x} (1-\mu) \right] \, \Gamma_{\rm free} 
\end{eqnarray}
to a very good approximation. Here, the value $\xi_{a,b}$ in Eq.~(\ref{decayratesimplified}) is a mirror parameter, which depends on the rate of reflection of either side of the mirror,
\begin{eqnarray}\label{mirrorparam}
\xi_{a,b} = \frac{3r_a r_b^2}{(r_a^2 + r_b^2)} \, .
\end{eqnarray}
As shown in Fig.~\ref{figxi}, this parameter varies between 0 and 1.5. Moreover, the parameter $k$ in Eq.~(\ref{decayratesimplified}) is the wave number of the emitted light and $\mu$ is the $x$-component of the normalised atomic transition dipole moment and varies between 0 and 1. 

\begin{figure}[ht]
	\begin{center}
		\begin{tabular}{c}
			\includegraphics[width=0.47 \textwidth]{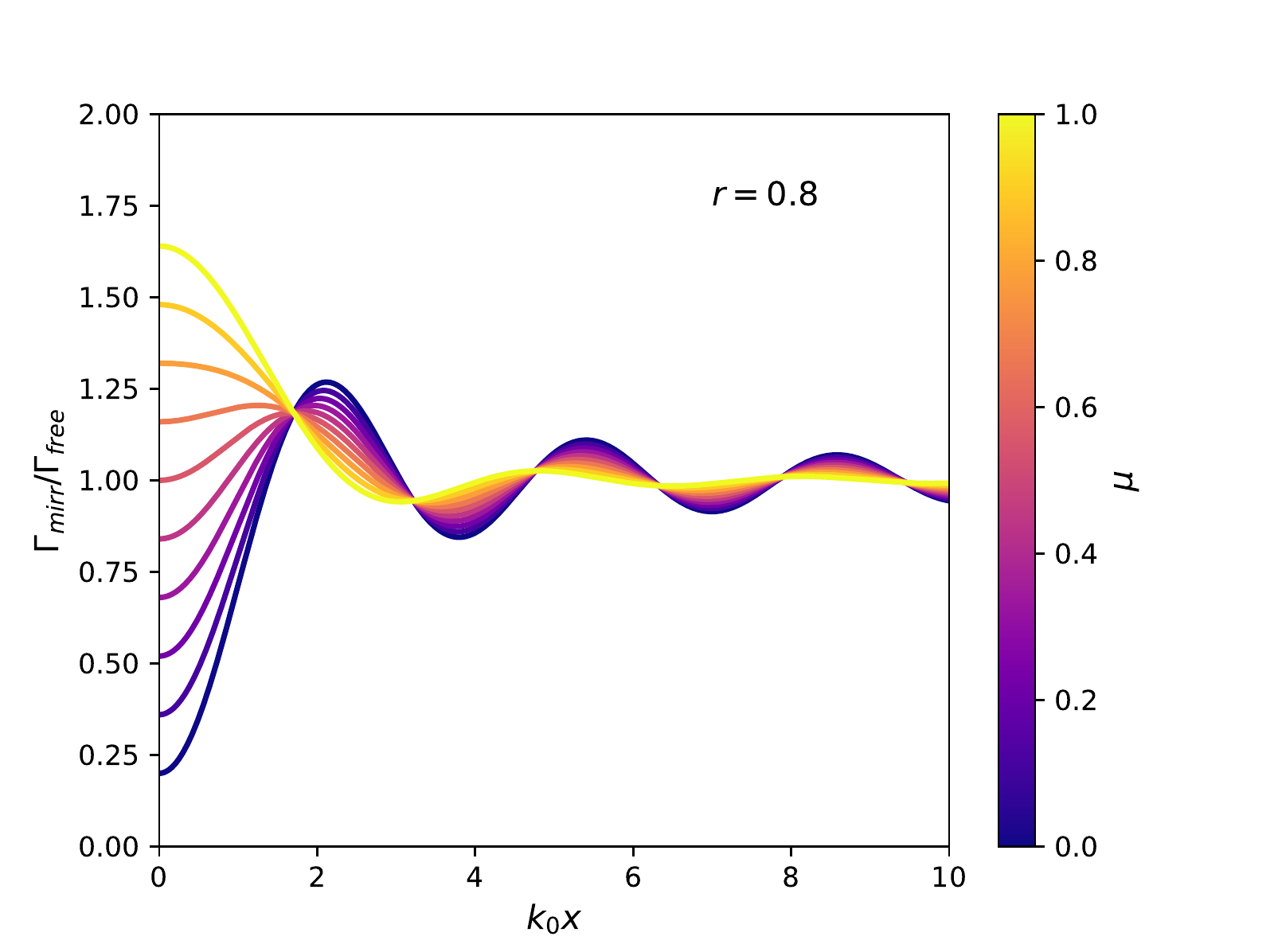}
		\end{tabular}
	\end{center}
	\caption[figmudep] 
	{\label{figmudep}
		The dependence of the spontaneous decay rate $\Gamma_{\rm mirr}(x)$ on the $x$ component of the transition dipole moment $\mu$ when $r_a=r_b=r$. For $\mu = 0$ and for $\mu = 1$, the atomic dipole moment is either parallel or orthogonal to the mirror surface. }
\end{figure}

Finally, Fig.~\ref{figmudep5} shows the dependence of $\Gamma_{\text{mirr}}(x)$ on the atom-mirror distance $x$ for different mirror parameters $\xi_{a,b}$ and for two different values of $\mu$.  It can be seen that the atomic lifetime has an oscillatory variation with distance. From Fig.~\ref{figmudep}, where the rate of reflectance $r_a=r_b=r$, we see that the amplitudes of these oscillations depend on the orientation $\mu$ of the atomic dipole moment. Our calculations confirm that the spontaneous decay rate $\Gamma_{\text{mirr}}(x)$ depends on both reflection rates, $r_a$ and $r_b$, of the mirror interface. Different from what one might naively expect, $\Gamma_{\text{mirr}}(x)$ so depends on the properties of both sides of the semi-transparent mirror interface. One way of explaining this effect is to interpret it as a dipole-dipole interaction between the atomic dipole and its mirror image (see e.g.~Refs. \cite{drabe, kuhn}). Consequently, a transition dipole perpendicular to the mirror will have a mirror transition dipole appearing to oscillate out of phase and quickly extinguishing each others influence over large distances. Conversely, transition dipoles parallel to the mirror will have a mirror transition dipole oscillating in phase, enhancing each others influence.

\section{Conclusions} \label{secconclusions}

This paper reviews the main ideas and results of a quantum mirror image detector method \cite{furtak,dawson} which can be used to model the quantised electromagnetic field in the presence of a two-sided semi-transparent mirror where the reflectance and transmittance can differ on either side. Here we use this method to predict the spontaneous decay rate of an atom in front of such a mirror (c.f.~Fig.~\ref{figpaperlogo}).  Although the effect which we describe here is relatively short-range, it can aid the design of novel quantum photonic devices. 

\acknowledgments 

We acknowledge financial support from the Oxford Quantum Technology Hub NQIT (grant number EP/M013243/1). Statement of compliance with EPSRC policy framework on research data: This publication is theoretical work that does not require supporting research data.

\end{document}